\documentclass{aa}
\usepackage[dvips]{graphicx}
\usepackage[]{txfonts}
\def\simless{\mathbin{\lower 3pt\hbox
{$\rlap{\raise 5pt\hbox{$\char'074$}}\mathchar"7218$}}}   
\def\simmore{\mathbin{\lower 3pt\hbox
{$\rlap{\raise 5pt\hbox{$\char'076$}}\mathchar"7218$}}}   
\newcommand{\be}{\begin{equation}}
\newcommand{\ee}{\end{equation}}

\begin{document}
\title{Spectral and timing properties of a dissipative GRB photosphere}
\titlerunning{Properties of a dissipative GRB photosphere}
\authorrunning{Giannios and Spruit}
\author{Dimitrios Giannios
\and Henk C. Spruit}

\institute{Max Planck Institute for Astrophysics, Box 1317, D-85741 Garching, Germany}

\offprints{giannios@mpa-garching.mpg.de}
\date{Received / Accepted}

\abstract
{We explore the observational appearance of the photosphere of 
an ultrarelativistic flow with internal dissipation of energy
as predicted by the magnetic reconnection model. Previous study 
of the radiative transfer in the photospheric r
egion has shown that 
gradual dissipation of energy results in a hot photosphere. There,
inverse Compton scattering of the thermal radiation advected with the flow
leads to powerful photospheric emission with spectral properties close 
to those of the observed prompt GRB emission. Here, we build on that study by 
calculating the spectra for a large range of the characteristics of the flow.
An accurate fitting formula is given that provides the photospheric 
spectral energy distribution in the $\sim 10$ keV to $\sim$10 MeV 
energy range (in the central engine frame) as a function of  the basic physical 
parameters of the flow.  It facilitates the direct comparison of the model 
predictions with observations, including the variability properties of the 
lightcurves. We verify that the model naturally accounts for the observed 
clustering in peak energies of the $E\cdot f(E)$ spectrum.  In this
model, the Amati relation indicates a tendency for the most luminous 
bursts to have more energy per baryon. If this tendency also holds
for individual GRB pulses, the model predicts the observed narrowing of the
width of pulses with increasing photon energy. }

\maketitle

\section{Introduction} 
\label{intro}

Despite the wealth of observational data obtained over the last few decades, the radiative
mechanisms responsible for the prompt emission of a $\gamma$-ray burst 
(hereafter GRB) remain elusive.  From observations of thousands of 
bursts it has been shown that the GRB prompt spectral energy distribution 
is characterized by a peak (in $E\cdot f(E)$ representation) that clusters 
in the sub-MeV energy range. The peak smoothly connects to low and high
frequency power-laws (Band et al. 1993; Preece et al. 1998; Kaneko et
al. 2006).  Theoretical arguments make use of the fact that a large 
fraction of the observed radiation is at photon energies above the pair 
creation threshold.  Together with the huge luminosities involved in 
the GRB phenomenon,  this implies that the emitting material is 
ultrarelativistic with bulk Lorentz factor $\Gamma\simmore 100$ (for a
review of these arguments see, for example, Piran 2005). Models for the prompt
GRB phase thus have to account for both the acceleration of the flow and its 
observed spectral properties.

For a flow to be accelerated to high bulk Lorentz factors, it must start with 
high energy-to-rest-mass ratio. Depending on whether the energy is in thermal 
or magnetic form, one has a fireball (Paczynski 1986; Goodman 1986) or a 
Poynting-flux dominated flow (Thompson 1994; M\'esz\'aros \& Rees 1997; Spruit 
et al. 2001; Drenkhahn \& Spruit 2002; Lyutikov \& Blandford 2003; Uzdensky \&
MacFadyen 2006). In the 
fireball model, the flow goes through an initial phase of rapid acceleration 
where $\Gamma\sim r$ until most of the internal energy has been used to 
accelerate the flow (unless the baryon loading is very low and radiation 
decouples from matter before the acceleration phase is over). Further out, 
internal shocks that are a result of fast shells catching up with slower ones 
(Rees \&  M\'esz\'aros 1994; Sari \& Piran 1997) can dissipate part of the
kinetic energy and power the prompt emission. In magnetic models for GRBs, the 
flow acceleration is more gradual (e.g. Drenkhahn \& Spruit 2002). 

Dissipation of magnetic energy through magnetic
reconnection (Drenkhahn 2002) or current driven instabilities (Lyutikov \& 
Blandford 2003; Giannios \& Spruit 2006) can directly power not only
the prompt emission, but also the conversion of Poynting flux to kinetic energy 
of outflow.  In fact,  at high Lorentz factors magnetic dissipation is  the only 
efficient mechanism for the conversion of Poynting flux to kinetic energy 
(Drenkhahn 2002, Drenkhahn \& Spruit 2002, Spruit \& Drenkhahn 2004;
for a disagreeing view see Vlahakis \& K\"onigl 2003). This  is discussed
again in section \ref{accel}.

Unlike the internal shock model, where dissipation of flow energy is due to 
the variability of the central engine, variability and dissipation are, a priori, 
unrelated in magnetic dissipation models; dissipation takes place even in 
a steady magnetic outflow.

If the dissipated energy leads to fast particles, then the synchrotron or
synchrotron-self-Compton mechanism appears as a natural one for the prompt 
emission. On the other hand, the large number of bursts with low energy slopes 
steeper than those that the synchrotron model can explain (e.g. Crider et al. 
1997; Frontera et al. 2000; Girlanda et al. 2003), have triggered alternative 
suggestions for the origin of the GRB emission. These include saturated 
Comptonization (Liang 1997) and Comptonization by thermal electrons
(Ghisellini \& Celotti 1999).

Another possibility for the origin of the prompt emission is the photospheric
emission of the flow (M\'esz\'aros \& Rees 2000; Ryde 2004; Thompson et
al. 2006). A common feature
of fireball and magnetic models is that they are characterized by (weaker or
stronger) emission from the region where the flow becomes optically thin
to Thomson scattering. If no energy dissipation takes place in that region, as 
in the case of fireball models, the emission is expected to be quasi-thermal
and a corresponding peak should be recognizable in the spectrum 
(Daigne \& Mochkovitch 2002), contrary to observation. On the other
hand, if there is substantial energy dissipation (through internal shocks, 
magnetic reconnection or MHD instabilities) close to the photospheric region,
the emitted spectrum can strongly deviate from a black body  (Rees \& 
M\'esz\'aros 2000; Pe'er et al. 2005; Ramirez-Ruiz 2005).

Giannios (2006, hereafter G06) studied the spectrum of the photospheric
emission in the magnetic reconnection model (Drenkhahn 2002; Drenkhahn 
\& Spruit 2002). The reconnection model makes definite predictions 
for the characteristics of the flow and the rate of magnetic energy 
dissipation at different radii,  enabling detailed study of the 
radiative transfer. This study showed that while radiation and 
matter are in approximate thermodynamic equilibrium deep inside the flow,
 direct heating of the electrons through energy dissipation leads to 
an increase of their temperature even at optical depths $\sim 50$. 
Inverse Compton scattering of the underlying thermal radiation leads to
 a spectrum with  a non-thermal appearance,  peaking at $\sim 1$ 
MeV (in the central engine frame) followed by a nearly flat high energy 
spectrum ($E\cdot f(E)\sim E^{0}$) extending to at least a few hundred MeV.

Here, we build on that study with a wide parametric exploration of 
the spectral energy distribution of the photospheric emission  in this 
model.  We find that the photospheric spectra, computed from 
detailed Monte-Carlo simulations, can be accurately modeled by a smoothly 
broken power law model in the 10-$10^4$ keV range.  These are then used 
to derive analytic expressions for the dependence of the  
spectra on the characteristics of the flow.  Since modulations of the 
characteristics of the flow by the central engine result in variable 
photospheric emission, modeling of the spectral and temporal properties
of the photospheric emission offers a unique probe into the operation of 
the central engine.
 
We look at the properties of the distribution of the peak energy of the
spectrum. The Amati relation is reproduced provided that the more luminous
flows are characterized by more energy per baryon (i.e. less baryon loading).
If the same baryon loading-luminosity corelation holds during the temporal
evolution of GRBs, the model predicts the observed energy dependence of the
width of the pulses of the GRB lightcurves.     

In the next section, we review the main features of the magnetic reconnection 
model and the treatment of the radiative transfer in the photospheric region. 
The analytical modeling of the photospheric spectra is given in Sect.~3. 
Inferences concerning the properties of the central engine are given
in Section 4 and we summarize our findings in the Discussion Section. 

\section{Dissipative GRB photosphere}
\label{model}

The GRB flow, whether magnetized or not, has a characteristic radius at which
it becomes transparent. This is  the radius where radiation and 
particles decouple and the photospheric emission originates. If 
no energy is dissipated close to the photosphere, the photospheric 
emission is expected to be quasi-thermal. On the other hand, significant 
dissipation of energy can strongly modify the photospheric spectrum. 
The physical source of energy dissipation can be the collision of shells of 
varying bulk Lorentz factors resulting in internal shocks in the flow (if, for 
some reason, the shells collide close to the photosphere). Dissipation near the 
photosphere is generic in the magnetic reconnection model, where the
dissipation of magnetic energy is gradual and takes place over several decades
of radius,  typically including the photosphere.

Unlike the internal shock model, where dissipation of flow energy is due to 
the variability of the central engine, variability and dissipation are, a priori, 
unrelated in the magnetic reconnection model; dissipation takes place even in 
a steady magnetic outflow. For typical inferred GRB parameters, the
dissipation in a magnetic reconnection model takes place around the
photosphere of the flow (Drenkhahn 2002). The time for the flow to reach the 
photosphere as observed by an on-axis external observer is so short that the 
flow can be treated as approximately steady for all but the shortest observed 
time scales. This greatly simplifies the interpretation of light curves and 
variations in the energy spectra. Of course, this applies only if it can be 
demonstrated that a substantial fraction of the emission in the 10-1000 keV 
range is in fact produced around the photosphere of the flow. The spectra 
obtained in G06 are promising in this respect.

The reconnection model makes specific predictions for the rate of energy 
dissipated as a function of radius; this makes it possible to study the radiative 
transfer in the flow in detail (G06). Analytic estimates demonstrate that the 
flow develops a hot photosphere where inverse Comptonization results in 
upscattering of the photons advected with the flow. The numerical (Monte
Carlo) calculations in G06 show that the spectra have many properties of the 
observed prompt GRB emission, for a wide range of baryon loading and
luminosities of the flow.

In the following subsections, we review the basic characteristics of the 
reconnection model and of the radiative transfer in the flow.

\subsection{The reconnection model}

An important physical quantity in the reconnection model is the ratio 
$\sigma_0$ of the Poynting flux to kinetic energy flux at the start of 
the flow, the Alfv\'en point $r_0$. This is a quantity equivalent to the 
`baryon loading' parameter in fireball models.  It determines the 
terminal bulk Lorentz factor of the flow $\Gamma_\infty\sim \sigma_0^{3/2}$. 
The flow must start Poynting-flux dominated with $\sigma_0\simmore 
30$ for it to be accelerated to ultrarelativistic speeds with $\Gamma_\infty 
\simmore 100$ that are relevant for GRB flows.

In the reconnection model, the magnetic field in the flow changes polarity on 
a scale $\lambda$. If the magnetic field anchored in the rotating central 
engine is nonaxisymmetric (the `AC model'), this scale is (in the central
engine frame) of the order of the light cylinder $r_{\rm l}$, ($\lambda\simeq 
r_0 \simeq 2\pi c/\Omega=2\pi r_{\rm l}$) where $\Omega$ is the angular 
frequency of the rotator. This is as in the oblique rotator model 
for pulsar winds (Coroniti 1990). The rate of magnetic reconnection in 
the model is parameterized through the velocity $v_{\rm rec}$ with which
magnetic fields of opposite direction merge. As in most models of magnetic 
reconnection, $v_{\rm rec}$ scales with the Alfv\'en speed, $v_{\rm A}$, 
$v_{\rm rec}=\varepsilon v_{\rm A}$. A nominal value used for $\epsilon$ 
is 0.1. For the flows with $\sigma_0\gg 1$  that are of interest 
here, the energy density of the magnetic field is larger than the rest mass 
energy density, hence $v_{\rm A}\approx c$, and the reconnection takes 
place with subrelativistic speeds. For a detailed analysis of 
relativistic reconnection see Lyubarsky (2005).

\subsubsection{Poynting flux conversion}\label{accel}
If the energy extracted from the central engine is initially in the form of a
Poynting flux, acceleration of the flow requires efficient conversion of this
flux into kinetic energy. The standard mechanism for acceleration of
nonrelativistic flows, magnetocentrifugal acceleration, (Blandford \& Payne 
1982) allows conversion of a substantial fraction (some 40\%), with the
rest remaining  in the flow as magnetic energy. For relativistic flows, the
conversion efficiency for this same mechanism is much  lower (Michel 1969),
so that in the case of a GRB almost all energy would end up in the
afterglow rather than as prompt emission.

Arbitrarily high effiency of conversion is possible, however, if there is a 
channel by which the magnetic energy carried by the flow can dissipate
by internal processes. Such dissipation results is an outward decrease
of magnetic pressure which accelerates the flow efficiently (Drenkhahn
2002; for a more detailed physical explanation of this at first sight somewhat
counterintuitive process see Spruit \& Drenkhahn 2004). 

The dissipation is gradual and takes place up to the `saturation radius' 
$r_{\rm s}$, the distance where reconnection stops and the flow achieves 
its terminal Lorentz factor. During the magnetic dissipation process, about 
half of the Poynting flux directly converts into kinetic flux (leading to 
acceleration) and the other half into internal  (thermal) energy of the 
flow. 

If the dissipation takes place under optically thin conditions, the 
thermal energy can be radiated away quickly. In this case half of the
Poynting flux is converted to radiation, the other half to kinetic energy of the
bulk flow. If radiation is inefficient, on the other hand, the internal energy 
adds to kinetic energy of the bulk flow, through adiabatic expansion (as in 
the case of a pure hydrodynamic fireball).

\subsubsection{Properties of the magnetically accelerated flow}
The reconnection model predicts gradual a acceleration of the flow 
$\Gamma\sim r^{1/3}$ in the regime $r_0\ll r\ll r_{\rm s}$, while no 
further acceleration takes place beyond the saturation radius. Summarizing 
the results derived in Drenkhahn (2002), the bulk Lorentz factor of the flow 
is approximately given by
\begin{eqnarray}
\Gamma&=&\Gamma_\infty\left(\frac{r}{r_{\rm s}}\right)^{1/3}=148~r_{11}^{1/3}(\varepsilon \Omega)_3^{1/3}
\sigma_{0,2}^{1/2}, \quad\rm{for}\quad r<r_{\rm s}, \nonumber\\
&&\label{gamma}\\
\Gamma&=&\Gamma_\infty=\sigma_0^{3/2}, \quad\rm{for}\quad r\ge r_{\rm s}, \nonumber
\end{eqnarray}
while the saturation radius is
\be
r_{\rm s}=\frac{\pi c \Gamma_\infty^2}{3 \varepsilon\Omega}; \quad{\rm or}\quad
r_{{\rm s},11}=310~\frac{\sigma_{0,2}^3}{(\varepsilon \Omega)_3}.
\label{rsatur}
\ee

In the steady spherical flow under consideration the comoving number density can be written as
\be n'=\frac{L}{r^2\sigma_0^{3/2}\Gamma m_{\rm p}c^3}, \label{ncom}\ee
where $L$ is the luminosity per steradian of the GRB model. (In this form the
expression can be compared with the fireball model, where the baryon loading
parameter $\eta$ replaces the factor $\sigma_0^{3/2}$). With eqs. (\ref{gamma}),
(\ref{ncom}) we have \begin{eqnarray}
n'&=&\frac{1.5\cdot 10^{17}}{r_{11}^{7/3}}\frac{L_{52}}{(\varepsilon \Omega)_3^{1/3}\sigma_{0,2}^2}\quad\rm{cm^{-3}}
\quad\rm{for}\quad r<r_s, \nonumber\\
&&\label{density}\\
n'&=&\frac{2.2\cdot 10^{19}}{r_{11}^2}\frac{L_{52}}{\sigma_{0,2}^3},
\quad\rm{cm^{-3}}\quad\rm{for}\quad r\ge r_s  \nonumber.
\end{eqnarray}
The comoving magnetic field strength B' is given by
\be
B'=\Big(\frac{4\pi L}{cr^2\Gamma^2}\Big)^{1/2}=\frac{1.4\cdot
  10^{8}}{r_{11}^{4/3}}\frac{L_{52}^{1/2}}
{(\varepsilon \Omega)_3^{1/3}\sigma_{0,2}^{1/2}}\quad\rm{G},
\quad\rm{for}\quad r<r_s 
\ee
with the magnetic field in the central engine frame given by $B=\Gamma B'$.
[The usual notation $A=10^xA_x$ is used; the `reference values' of the
model parameters are $\sigma_0=100$, $\varepsilon=0.1$, $\Omega=10^4$
rad$\cdot$s$^{-1}$, $L=10^{52}$ erg$\cdot$s$^{-1}\cdot$sterad$^{-1}$.]

\subsection{Radiative transfer in the photospheric region}

In addition to the saturation radius $r_{\rm s}$,
another characteristic radius of the flow is the Thomson photosphere.
 A photon moving radially outward in a flow with $\Gamma \gg 1$ encounters an
 optical depth $d\tau=n' \sigma_{\rm T} {\rm d}r/2 \Gamma$ (Abramowicz et
 al. 1991). Integrating from $r$ to $\infty$ the characteristic 
Thomson optical depth as a function of radius is
\be
\tau=\frac{20}{r_{11}^{5/3}}\frac{L_{52}}{(\varepsilon \Omega)_3^{2/3}
\sigma_{0,2}^{5/2}}.
\label{tau}
\ee
Setting $\tau=1$ yields the photospheric radius $r_{\rm ph}$
\be
r_{\rm{ph,11}}=6\frac{L_{52}^{3/5}}{(\varepsilon \Omega)_3^{2/5}
\sigma_{0,2}^{3/2}}.
\label{rphot}
\ee
In deriving these expressions, we have assumed that  $r_{\rm{ph}}<r_s$.
At high baryon loading, the Lorentz factors can become so low that most 
of the Poynting flux dissipates already inside the photosphere, i.e. 
$r_{\rm{ph}}>r_s$. With the expression for the number density valid 
beyond the saturation radius from (\ref{density}), a similar 
calculation gives the radius of the photosphere in this case.

Due to the dominance of scattering, the details of radiative transfer 
become important already at fairly large optical depth in the flow. 
Equilibrium between radiation and matter holds only at Thomson depths 
greater than about 50. At optical depths $0.1\simless\tau\simless 50$ 
the electron distribution stays thermalized, but is out 
of equilibrium with the photon field. Compton scattering of the photons 
needs to be treated in detail in this region. This has been done in G06, 
using both analytical and numerical tools. One of the main findings 
of this study is that the nature of the photospheric spectrum is determined by 
the location of the photosphere relative to the saturation radius. If 
$r_{\rm{ph}}\gg r_s$, all the energy dissipation takes place in optically 
thick conditions, the radiation is efficiently thermalized and suffers 
substantial adiabatic losses. On the other hand, if $r_s\simmore r_{\rm{ph}}$,
 energy dissipation at moderate and low optical depths leads to a photospheric 
emission that has a highly non-thermal appearance. 
 
There is thus a characteristic spectral transition of the photospheric
emission depending on the relative locations of the photospheric and
saturation radii. In terms of the 
physical properties of the flow, it takes places at the critical value 
of the baryon loading  $\sigma_{\rm {0,cr}}$, for which 
$r_{\rm ph}=r_{\rm s}$. For $\sigma>\sigma_{\rm {0,cr}}$, 
$r_{\rm ph}<r_{\rm s}$. Using eqs.~(\ref{rphot}) and (\ref{rsatur}) one finds
\be
\sigma_{\rm {0,cr}}=42 \big(L_{52}(\varepsilon \Omega)_3\big)^{2/15}.
\label{sigmacr}
\ee

At radii much shorter than the photospheric radius, the released energy is 
efficiently thermalized and shared between particles and radiation. In this 
region, the comoving temperature $T_{\rm{th}}$ of the flow is calculated, 
under the assumption of complete thermalization, by integrating the energy 
released at different radii in the flow and taking into account adiabatic 
cooling. This leads to (Giannios \& Spruit 2005)
\be
T_{\rm{th}}=\frac{0.7}{r_{11}^{7/12}}\frac{L_{52}^{1/4}}{(\varepsilon \Omega)_3^{1/12} 
\sigma_{0,2}^{1/2}} \quad \rm{keV}.
\label{Tth}
\ee   

Complete thermalization of the dissipated energy is achieved up to a radius 
defined as the {\it equilibrium radius} 
\be
r_{\rm{eq,11}}=0.6\frac{L_{52}^{5/9}}{f_{\rm e}^{4/9}(\varepsilon \Omega)_3^{1/3}\sigma_{0,2}^{4/3}},
\label{req}
\ee  
where $f_{\rm e}$ stands for the fraction of the dissipated energy 
that heats the electrons  (rather than the ions) and is assumed to 
be of order of unity. As a reference value we adopt $f_{\rm e}=0.5$.
Note that the equilibrium radius lies at a distance 
one order of magnitude below the Thomson photosphere. At radii
$r>r_{\rm{eq}}$, direct electron heating through magnetic dissipation 
leads to electrons with characteristic energy  higher than that of 
the radiation field and upscattering of the photons takes place in this 
region. Because of the rather steep increase of the electron temperature 
 with radius above $r_{\rm eq}$, this upscattering becomes efficient 
 especially around the location of the photosphere, where the comoving 
electron temperature is of the order a few tens keV.  For these
temperatures  and optical depths the result is a hard power law tail in the 
spectrum, characteristic of unsaturated Comptonization.

 For the detailed calculation of the emergent spectrum and the electron 
temperature above the equilibrium radius we use a Monte Carlo method as 
described in G06. It treats Compton scattering in a medium with density and 
bulk Lorentz factor given by eqs. (\ref{density}) and (\ref{gamma})
respectively. The special relativistic effects 
related to both the bulk motion and the scattering cross-section of photons 
in the flow are taken into account. The inner boundary of the computational 
domain is the equilibrium radius defined by eq. (\ref{req}). There, a black 
body photon spectrum is injected with temperature given by eq. (\ref{Tth}). 
The outer boundary $r_{{\rm out}}$ is, rather more arbitrarily taken  at a 
radius where the optical depth becomes small (the value $\tau=0.1$ is 
used for the rest of the Section, but see Section (\ref{tauout}) where the 
dependence of the emitted spectrum on the choice of $r_{{\rm out}}$ is investigated).

The electron temperature as a function of distance in the flow is determined 
from the balance of radiative cooling by Comptonization and heating of the 
electrons by magnetic dissipation. The numerical results verify the analytical 
expectation that the electron temperature increases rapidly in the
photospheric region and that significant Comptonization takes place there (G06).

\begin{figure}
\resizebox{\hsize}{!}{\includegraphics[angle=270]{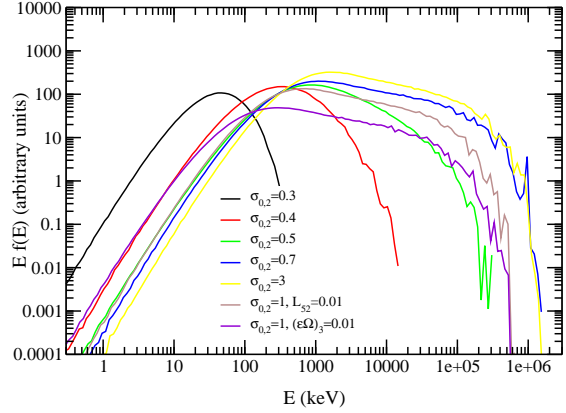}}
\caption[]
{Photospheric spectrum of the 
reconnection model for GRB flows (photon energies in the central engine frame). 
For moderate and high initial Poynting flux ratio $\sigma_0$ the spectra peak
around 1 MeV. The flat high energy tail is the result
of inverse Compton scattering of photons off hot electrons close to the
photospheric radius. For low $\sigma_0$ dissipation stops below the
photosphere, and the resulting spectrum is quasi-thermal. 
\label{fig1}}
\end{figure}

A selection of characteristic emergent spectra from the calculations are 
presented in Fig.~\ref{fig1} in $E\cdot f(E)$ representation for different 
values of the parameters of the model. The spectra are plotted in the central 
engine frame with arbitrary normalization (the energetics are discussed in 
the next Section). The peak of the spectrum is close to $~1$ MeV for a large 
region of the parameter space which is relevant for GRB flows. The dependence 
of the peak energy on the parameters of the model is very weak. This can be 
understood from the fact that the peak energy is related to the peak of the 
thermal emission advected with the flow at the equilibrium radius which, in 
the central engine frame, is $E_{\rm p, th}\propto T_{\rm eq}^{\rm ce}\propto 
\Gamma(r_{\rm {eq}})T_{\rm th}(r_{\rm eq})\propto L_{52}^{1/9}(\varepsilon 
\Omega)_3^{1/3}\sigma_{0,2}^{1/3}$, where in the last step the expressions 
(\ref{Tth}), (\ref{gamma}) and (\ref{req}) have been used. The detailed
numerical investigation leads to very similar dependence of $E_{\rm p}$ on 
the model parameters (see eq.~\ref{Ep}).

The inverse Compton scattering that takes place above the equilibrium radius 
leads to a moderate increase of the energy of the peak and, most important, 
to a power-law high energy emission with photon number index $\sim -2.5$ that 
extends up to a few hundred MeV. This high energy part of the spectrum is the 
result of unsaturated Comptonization taking in the $\tau \simless 1$ region 
and appears for flows with $\sigma_0\simmore \sigma_{\rm 0,cr}$.

In the lower $\sigma_0$ cases, the emergent spectrum has much weaker emission 
above its peak. In these cases the magnetic dissipation stops below the
Thomson photosphere and there is only weak Compton upscattering taking place 
in the photospheric region.  Such quasi-thermal emission has been
observed in a fraction of GRBs (Ryde 2004; 2005)  suggesting these may 
be bursts of low $\sigma_0$ (high baryon loading).

\section{Analytical fits and parametric study of the spectra}  

The spectra obtained with our Monte-Carlo radiative transfer calculations 
have systematic dependencies on the parameters of the model (initial Poynting
flux to kinetic flux ratio $\sigma_0$, luminosity per steradian $L$ and reconnection rate 
which appears in the combination $\epsilon \Omega$). In this section we
condense these dependences into a more practically useful form by developing 
fitting formulas.

\begin{figure}
\resizebox{\hsize}{!}{\includegraphics[angle=270]{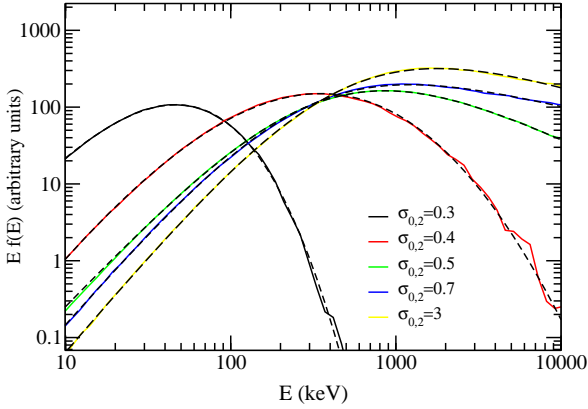}}
\caption[]
{Smoothly-broken power law fits (dashed) to the spectra of Fig. \ref{fig1} 
(solid). Photon energies in the central engine frame. The 5-parameter SBPL 
formula (eq.~(\ref{SBPL})) fits the numerically calculated spectra very
accurately in the 10 keV to 10 MeV energy range.  
\label{fig2}}
\end{figure}

After various attempts to find analytical functions that fit the numerically 
computed spectra, it turned out that the  smoothly 
connected power law (hereafter SBPL model Preece et al. 1994; Ryde 1999;
Kaneko et al. 2006) provides a very 
good description in the 10 keV to 10 MeV energy range,  which is the most 
relevant for observations of the prompt emission of GRBs. The SBPL is a 
generalization of the  sharply broken power 
law model by the addition of a parameter governing the width of the 
transition from the low energy spectral slope to the high energy slope. The
fitting model has five parameters: an amplitude $A$, low and high photon indexes 
$\lambda_1$ and $\lambda_2$, a break energy $E_{\rm{b}}$ and a break scale 
$\Lambda$.  The transition in slope is described by a hyperbolic tangent function.

The SBPL model fits the data (in our case the spectrum) with the
expression\footnote
{We choose a representation of the SBPL model where only the natural 
logarithm appears. This is achieved by suitably renormalizing the break scale with
respect to that used by Ryde (1999).}
(for the derivation see for example Ryde 1999) 
\be
f(E)=A\Big(\frac{E}{100}\Big)^{\frac{\lambda_1+\lambda_2+2}{2}}\Big( 
\frac{e^q+e^{-q}}{e^{q_{\rm piv}}+e^{-q_{\rm
      piv}}}\Big)^{\frac{\lambda_2-\lambda_1}
{2}\Lambda},\label{SBPL}
\ee
where
\be
q=\frac{\ln(E/E_{\rm b})}{\Lambda}, \quad 
q_{\rm piv}=\frac{\ln(100/E_{\rm b})}{\Lambda} \nonumber.
\ee
The amplitude $A$ is in photons $s^{-1}$  and all energies in keV.

The break energy $E_{\rm b}$ is related to the peak energy $E_{\rm p}$ of the  
distribution $Ef(E)$ by
\be E_{\rm p}=E_{\rm b}\Big[-{\lambda_1+2\over\lambda_2+2}\Big]^{\Lambda/2}\label{EpEb}\ee
(e.g., Kaneko et al. 2006)\footnote{The very existence of a peak in the
spectrum demands that $\lambda_1+2>0$ and $\lambda_2+2<0$, which almost always
holds here.}. 
  
Since this fitting model does not have a high energy exponential break, 
it cannot describe the spectra above $\sim 100$ MeV. The SBPL fits also show 
small systematic deviations for energies below $\sim$several keV. Thus, we 
limit our fits to the, nevertheless broad, 10 KeV - 10 MeV energy range. We
have used the SBPL model to fit a large number of theoretical spectra for 
models of varying luminosity $L$, baryon loading $\sigma_0$ and reconnection 
rate parameterized by $\varepsilon \Omega$.

A few, rather representative, examples of such fits are shown in
Fig.~\ref{fig2}. The fits (shown as dashed curves) lie very close to 
the numerical results with the exception of the lowest end of the 
energy range considered. Overall, SBPL fits are a very good 
way to summarize the photospheric spectrum predicted by the dissipation model.

With these fits, the low and high energy slopes $\lambda_1$ and $\lambda_2$ 
and the peak energy $E_{\rm p}$ of the spectrum can be presented as functions 
of the parameters of $L$, $\sigma_0$ and $\varepsilon \Omega$ of the model.

\begin{figure}
\resizebox{\hsize}{!}{\includegraphics[angle=270]{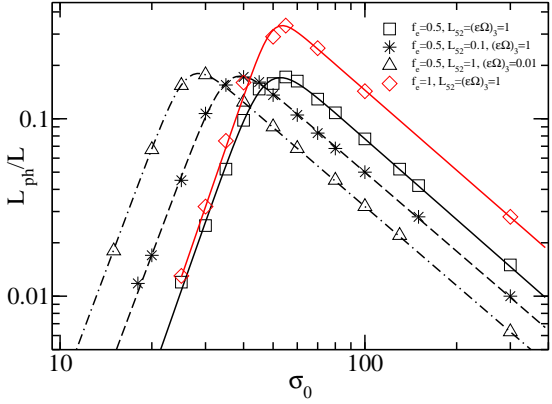}}
\caption[]
{ The (photospheric) radiative efficiency of the flow (ratio of photospheric 
luminosity to flow luminosity), as a function of the initial Poynting flux to 
kinetic energy flux ratio $\sigma_0$. At low $\sigma_0$, most of the 
dissipation happens at great optical depth, and the photospheric radiation 
field is reduced by adiabatic expansion. At high $\sigma_0$ most dissipation 
takes places beyond the photosphere. The solid red line shows the effect of 
increasing the electron fraction of the dissipated energy from 50 to 100\%.
Changes in luminosity $L$ and $\varepsilon \Omega$ are equivalent to shifts 
along the $\sigma$-axis. The curves are fits to smoothly broken power laws
(best fit values in the appendix A.2).
\label{fig3}}
\end{figure}

\begin{figure}
\resizebox{\hsize}{!}{\includegraphics[angle=270]{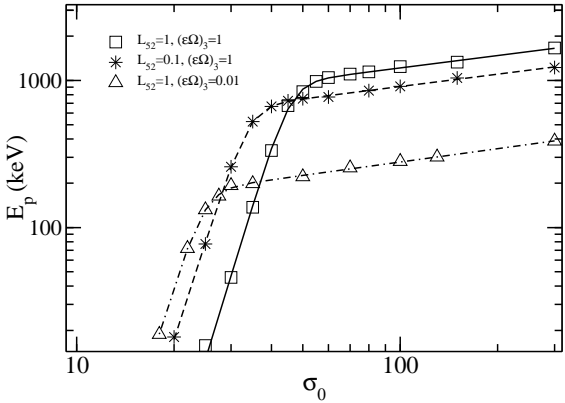}}
\caption[]
{Dependence of peak energy (peak of the photospheric $E\cdot f(E)$ spectrum) 
on baryon loading, flow luminosity and reconnection speed. Above the 
characteristic Poynting flux parameter $\sigma_{0,{\rm cr}}$ (Eq. 7), $E_{\rm p}$ 
depends only weakly on baryon loading. At high baryon loading (low $\sigma_0$), 
the spectrum becomes quasithermal (cf. Fig. 1), peaking at a low energy. Curves 
show the smoothly broken power law fits (cf. appendix A.2). 
\label{fig4}}
\end{figure}

Fig.~\ref{fig3} shows the ratio of the photospheric emission to total 
flow luminosity, i.e. the (photospheric) radiative efficiency of the model
and Fig.~\ref{fig4} the peak energy $E_{\rm p}$ of the spectrum.
For $\sigma_0$ around the critical value $\sigma_{\rm 0,cr}$, the radiative
efficiency peaks at $\approx 0.35f_{\rm e}$, {\it independent} of the luminosity of the 
flow $L$ and the reconnection rate parameter $\varepsilon \Omega$ . For 
$\sigma_0<\sigma_{\rm
 0,cr}$, the magnetic dissipation stops below the photosphere and radiation 
suffers substantial adiabatic cooling before it decouples from matter; this 
results in a steep increase of the radiative efficiency and of the peak energy
as functions of $\sigma_{\rm 0}$. Changes in $L$ and $\varepsilon \Omega$ 
lead to characteristic displacements of the curves in the $L_{\rm
ph}/L-\sigma_0$  and $E_{\rm p}-\sigma_0$ planes (see Figs.~3 and 4).

Analytic fits for the radiative efficiency and peak energy as functions 
of the characteristics of the flow are given in the appendix. These expressions 
 reduce to simple power laws when $\sigma\simmore \sigma_{\rm cr}$. 
Thus, for a sufficiantly high initial magnetization parameter in the flow, the
radiative efficiency 
has the following dependence on the model parameters (see Sect. A.2)
\be L_{\rm ph}/L\approx 0.07
\sigma_{0,2}^{-1.5}(\varepsilon \Omega)_3^{0.2}L_{52}^{0.2},\quad \sigma_0>
\sigma_{\rm 0,cr}. \label{effic} \ee
Similarly, for $\sigma_0>\sigma_{\rm 0,cr}$ the expression for the peak energy 
simplifies to (see Sect. A.3)
\be
E_{\rm p}\approx 1200L_{52}^{0.11}(\varepsilon \Omega)_3^{0.33}\sigma_{0,2}^{0.3}
\quad {\rm keV},\quad \sigma_0> \sigma_{\rm 0,cr}.
\label{Ep}
\ee 
 
The high-energy photon index $\lambda_2$ is almost constant at $\sim-2.5$ 
for $\sigma_0\simmore \sigma_{\rm 0,cr}$ and decreases rapidly for smaller
$\sigma_0$ (because there is little or no dissipation close to the photosphere 
for low values of $\sigma_0$). The  expression for $\lambda_2$ as a
function of the characteristics of the flow is given in the appendix A.4.
The low-energy photon index $\lambda_1$ shows very small variations and
remains approximately $\sim 0$ in most the parameter space investigated.
For simplicity, we keep $\lambda_1=0$ independently of the
characteristics of the flow. On the other hand, significant softening of
$\lambda_1$ appears when the spectral peak approaches the lowest energy range 
over which the spectral fits are performed (i.e. at 10 keV).
This may lead to deviations of the fitting formulas and the actual predictions
of the photospheric model in the $\sim 10$ keV range (in the frame of the
central engine). The break scale $\Lambda$ is practically constant $\sim 1.6$
for high $\sigma_0$ and makes a transition to the value $\sim 2.8$ for
$\sigma_0<\sigma_{\rm 0,cr}$. This transition can be modeled by a hyperbolic 
tangent (see appendix A.2).

\subsection{The non-photospheric emission}
\label{tauout}

The spectra above were obtained with an outer boundary set at a scattering
optical depth of 0.1. In the optically thin part of the flow ($\tau<0.1$) 
the electron temperature will continue to rise if significant dissipation
takes place there, and this will contribute to the Comptonization. We have
not included this contribution so far because the assumption of a thermalized
electron distribution becomes questionable at $\tau\simless 0.1$, and a nonthermal
distribution will change the Comptonization process. This makes the emission 
in these regions dependent on poorly known details of the magnetic dissipation 
process.

\begin{figure}
\resizebox{\hsize}{!}{\includegraphics[angle=270]{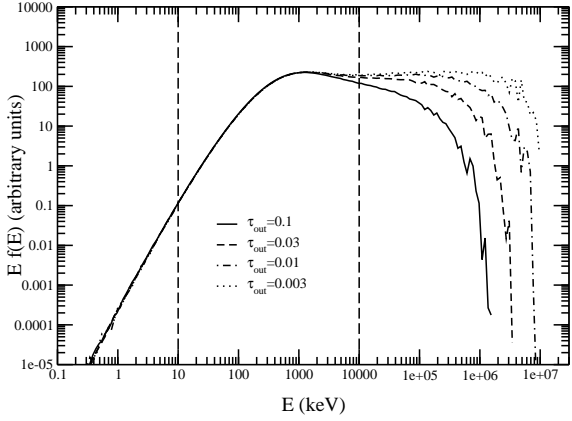}}
\caption[]
{Modification of the spectrum when an estimate of the emission in the 
optically thin part of the flow is included. [Electron distribution is assumed 
to remain thermal at low optical depths.] 
Parameter is the optical depth at the outer boundary of the calculation
(photon energies in the central engine frame).
Dashed lines mark the range used for the fitting formulas.

\label{fig5}}
\end{figure}

Notwithstanding this, it is possible to get an idea concerning at which
energy band the optically thin
emission may be expected to take place by extending the outer boundary
of the Monte Carlo calculation to larger radii (or equivalently to 
smaller optical depths in the flow). In this excercise we assume that 
the electrons still follow a thermal distribution in this region.
In Fig.~(\ref{fig5}) the emitted spectrum for the reference
values of the parameters of the model is shown for various values
of the optical depth $\tau_{\rm out}$ at the outer boundary of the 
numerical calculation.  The spectrum below the peak energy remains
essentially the same. The higher energy emission is significantly
enhanced as a result of scattering in the very hot, optically thin part 
of the flow. The high-energy spectral slope hardens, approaching 
$\lambda_2\sim-2$, and there is substantial emission in the $\sim$GeV 
energy range.  
  
The vertical dashed lines show the energy range used for the fitting
formulas presented in the previous section. The optically thin emission 
affects the high-energy spectral slope $\lambda_2$ somewhat, which  now 
lies in the $\sim -2$ to $-2.5$ range. 
The optically thin emission increases the luminosity in the 10-1000keV 
range only moderately, but can produce a significant peak in the GeV range.

\section{Interpretation of the observations}

In the following we use a first comparison with observations to make 
inferences about the properties of GRB outflows in the magnetic dissipation 
model.

\subsection{The value of $E_{\rm p}$}

The {\it observed} spectral peak of the prompt emission spectrum of GRBs  has
a narrow distribution peaking at $\sim$250 keV (Band et al. 1993; Kaneko et
al. 2006); by far the largest amount of data emanates from BATSE.  To compare 
this distribution with
the prediction of the photospheric model (\ref{Ep})  the redshift distribution 
of GRBs has to be taken into account. This distribution is not well known for the 
BATSE bursts, but various estimates indicate that an average value of $z\sim1.5-2$ 
is reasonable. For a typical burst at this redshift the model predicts that (cf. 
eq.~(\ref{Ep}))
\be 
 E_{\rm p}^{\rm obs}=\frac{E_{\rm p}}{1+z}\simeq 450L_{52}^{0.11}(\varepsilon
 \Omega)_3^{0.33}\sigma_{0,2}^{0.3} \quad {\rm keV}.
\label{Epobs}
\ee

The dependence on the characteristics of the flow is weak; the model thus
predicts a narrow distribution of the peak energy of bursts, in agreement with 
observation. For the reference values of the parameters used in eq.~(\ref{Epobs}), 
the peak energy predicted by the model is at 450 keV, a factor of $\sim2$ 
above the value for BATSE bursts. Though this is not a large factor, the weak
dependence on model parameters implies a significant difference in one or more of 
them compared with the reference value. A low value for $\sigma_0\ll 100$ is not
attractive, since then the flow becomes too slow to explain the typical GRB. 
The very weak dependence of the peak energy on the luminosity also excludes the 
burst luminosity. Since the central engine is most likely a stellar mass compact
object, the rotation rate, $\Omega$, must be of the order $10^3-10^4$ rad s$^{-1}$. The most 
uncertain parameter of the magnetic dissipation model is actually the reconnection 
speed parameter $\varepsilon$, for which we have so far  assumed a value $\sim 0.1$ 
(Drenkhahn 2002). $\Omega=10^4$ rad s$^{-1}$ is perhaps on the high side for the torus in a 
collapsar, and $\varepsilon=0.1$ possibly too optimistic. A value $(\varepsilon 
\Omega)\sim 100$ rad s$^{-1}$ instead of $10^3$, which would bring the predicted 
value of $E_{\rm p}$ into agreement with observation, can easily be accommodated 
within these uncertainties. 

This indicates that magnetic dissipation
timescale is one order of magnitude longer than that assumed by the reference
values of the parameters. 
In the following we adopt $(\varepsilon \Omega)=10^2(\varepsilon \Omega)_2$
as a reference value for the parameterization of the dissipation timescale.
 
\subsection{The Amati relation}

There are indications in the literature that the peak energy $E_{\rm p}$ of a 
burst correlates with the isotropic equivalent energy: $E_{\rm p}\propto E_{\rm
iso}^{0.5}$ (Amati et al. 2002; Amati 2006), although this correlation is probably 
not universal (Nakar\& Piran 2005; Band \& Preece 2005). The isotropic equivalent 
energy of the prompt emission is $E_{\rm iso} \sim L_{\rm ph}t_{\rm GRB}$, where 
$t_{\rm GRB}$ is the duration of the burst. For an estimate of the 
energy of the prompt emission, the relevant luminosity is that of the photosphere
$L_{\rm ph}$ and {\it not} that of the flow  $L$.     

Using the expressions (\ref{Ep}) and (\ref{effic}) one can find the dependence  
on the parameters of the flow of the ratio
\be
\frac{E_{\rm iso}^{1/2}}{E_{\rm p}}\propto \frac{L^{0.5}t_{\rm GRB}^{0.5}}
{(\varepsilon \Omega)^{0.23}\sigma_0^{1.1}}.
\label{ratio}
\ee 
According to the Amati relation, this ratio stays approximately constant
over a wide range of $E_{\rm iso}$. Since the dispersion in the distribution
of durations of $t_{\rm GRB}^{1/2}$ for long GRBs is rather small 
$\sim 10^{1/2}$, it is unlikely that a dependence in a systematic way 
of $t_{\rm GRB}$ with another parameter (say the luminosity) is responsible
for keeping the ratio (\ref{ratio}) constant. On the other hand,    
the ratio is approximately constant for $\sigma_0 \propto L^{\sim
 0.4-0.5}$ in different bursts\footnote{Assuming that  $(\varepsilon \Omega)$ does not change much
from burst to burst; if it varies in a systematic way with, say, the
luminosity, the  scaling of $\sigma_0$ with $L$ can be somewhat different.}.
The Amati relation, therefore, indicates that there is a corelation
of baryon loading and luminosity, in the sense that the most luminous
GRB flows have more energy per baryon. We see in the next Section that if 
the same baryon loading-luminosity corelation holds during the 
evolution of individual GRBs, the model predicts the observed energy
dependence of the width of pulses in the GRB lightcurves.

\subsection{GRB variability and spectral evolution}

Up this point, we have dealt with a steady flow that is characterized by
fixed luminosity and baryon loading. This approach can, to some extent, 
describe the rough, time average, properties of a GRB. On the other hand,
the GRB lightcurves are highly variable and are composed by a sequence of 
pulses believed to be the ``building blocks'' of their lightcurves
(e.g. Fenimore et al. 1995).
The question that rises as to where the variability comes from in the
photospheric model for the prompt emission and under what conditions a
steady state model such as the magnetic reconnection one can describe a
variable flow.

An ultrarelativistic flow that varies on a 
timescale $\delta t$ can be still described by a steady state model as long
as $\delta t>r/\Gamma^2 c$. When the last condition holds, the leading and
trailing parts of a piece of the flow with width $\delta r \simeq c\delta
t$ have not come into causal contact.
As long as this is the case, different parts of the flow behave as part of a  
steady wind. A robust proof of this statement can be found in Piran et al. (1993)
for fireballs and Vlahakis \& K\"onigl (2003) for MHD flows. 
The relevant radius of interest in this study is the photospheric radius.
As far as the photospheric emission is concerned, the steady 
state approximation is valid for a flow which
varies on timescales
\be                         
\delta t>\frac{r_{\rm ph}}{\Gamma^2(r_{\rm ph})c}\simeq 2\cdot
10^{-3}\frac{L_{52}^{1/5}}{(\varepsilon
  \Omega)_2^{4/5}\sigma_{0,2}^{3/2}}\quad {\rm s},
\ee
where eqs.~(\ref{rphot}) and (\ref{gamma}) are used in the last step.
The flow can treated as approximately steady for all but the shortest observed
(sub-millisecond) timescales.

\subsubsection{Central engine: the stronger the cleaner}

Variability and dissipation are, a priori, unrelated in the magnetic
reconnection model in which dissipation takes place, even in a steady magnetic
outflow. On the other hand, the flow does evidently evolve during a GRB.
In the context of the present model, the observed variability reflects changes in the
luminosity (and possibly the initial magnetization parameter or even 
the reconnection rate $\varepsilon \Omega$) during the burst. 
Since the flow up to the photosphere can be treated as
quasi-stationary for all but the shortest time scales observed in
a burst, the variation of spectral properties during a burst
directly reflects variations in the central engine. This is in
contrast to models in which the prompt radiation is produced at
much larger distances from the source, such as external shock
models. It is also in contrast with the internal shock model,
since the internal evolution of the flow between the source and
the level where radiation is emitted is a key ingredient in this
model. Deducing properties of the central engine from observed
burst properties is thus a much more direct prospect in the
magnetic dissipation model.

\begin{figure}
\resizebox{\hsize}{!}{\includegraphics[angle=270]{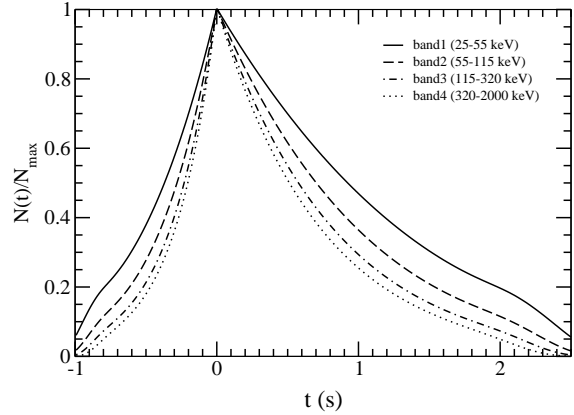}}
\caption[]
{Narrowing of pulse width with photon energy, as predicted by the
model for individual subpulses of the burst. The assumption is that the
relation between luminosity and baryon loading deduced from the
Amati relation, $\sigma_0\propto L^{0.4}$, holds also within a
burst.
\label{fig6}}
\end{figure}

As a first application we consider here the width of subpulses in
a burst. The observations show this width to be a decreasing
function of photon energy (e.g. Fenimore et al. 1995). 
GRBs with complex light curves (composed of many subpulses) follow 
the same Amati relation as simple ones (with one or a few pulses). 
In the above, we have shown how this relation implies a dependence of the baryon loading
of the burst on its luminosity. We now make the logical
extrapolation of assuming that this same relation also holds for
the time-dependent flow {\em during} a burst. Varying the luminosity
of the flow with time in such a way that it mimics the typical
pulse profile, one can simulate the pulse appearance predicted
by the photospheric model in different energy bands.  
To reproduce the characteristic pulse shape, we consider time dependence of the 
luminosity of the flow of the form (cf. Norris et al. 1996)
\begin{eqnarray}
L(t)&=&L_{\rm max}e^{-|t/t_{\rm r}|},\quad t<0\nonumber\\
&&\label{lum}\\ 
L(t)&=&L_{\rm max}e^{-t/t_{\rm d}},\quad t>0  \nonumber
\end{eqnarray}
where $t_{\rm r}=0.2/(1+z)$ s and $t_{\rm d}=0.5/(1+z)$ s are the rise and decay timescales of the
lightcurve (in the central engine frame) respectively and $L_{\rm
  max}=10^{52}$ erg s$^{-1}$ sterad$^{-1}$.  
We also assume that the baryon loading tracks the luminosity with the
expression  $\sigma_{0,2}=L_{52}^{0.4}$ (indicated by the Amati relation)
and set $\varepsilon \Omega=100$ rad s$^{-1}$. The redshift of the burst, needed
to correct for photon reshifting and cosmological time dilation, is taken to
be $z=2$. 

Having specified the parameteres of the
flow as functions of time, one can calculate the lightcurve (in the 
observer's frame) of the photopsheric emission at different energy bands 
predicted by the photospheric model using eq.~(\ref{SBPL}) and the expressions
in the appendix. In fig.~\ref{fig6}, the predicted lightcurves in the BATSE energy 
bands are plotted. The lightcurves of the harder energy bands are narrower 
than those of the softer. The width in the 320-2000 keV band being approximately half
of that of the 25-55 keV band close to what is observed. The energy dependence
of the lightcurves comes from the spectral hardening of the photospheric
emission with increasing luminosity $L$ (and therefore increasing photospheric
luminosity $L_{\rm ph}$; see eqs.~(\ref{effic})
and (\ref{Ep})). This spectral hardening results in relatively enhanced emission 
in the hard bands (with respect to the softer ones) when the luminosity is close to its peak.

\section{Discussion/conclusions}
\label{conclude}

In the magnetic reconnection model, the flow that powers
a GRB is initially Poynting flux dominated.
The model is characterized by gradual dissipation of
magnetic energy that leads to acceleration of the flow and an increase in its
internal energy. For characteristics of the flow typical to those expected
for GRB flows (i.e., high luminosities and low baryon loadings) the dissipation  
continues at distances outside the point where the flow becomes
transparent with respect to Thomson scattering (i.e. outside the Thomson photosphere).
The energy release close to the photospheric radius leads to a hot photosphere.
Inverse Compton scattering of the advected with the flow  photon
field by hot electrons leads to photospheric
spectra with a characteristic non-thermal appearance and with peak energy and 
spectral slopes close to those observed in the prompt phase of GRBs. The
radiative  transfer problem has been studied in detail, with both analytical
and numerical tools, in G06. 

Here, we build further on that study by providing an accurate analytical description of
the spectra predicted by the model in the x-ray and $\gamma$-ray regime. 
We show that the photospheric emission
from the reconnection model is well described by a smoothly broken power law
model. Simple fitting formulas are presented for the bolometric luminosity 
and peak energy of the $E\cdot f(E)$ spectrum as a function of  the flow
parameters (i.e. the luminosity of the flow $L$, the baryon loading $\sigma_0$ and
the reconnection timescale parameter $\varepsilon\Omega$). 
 
The model predicts the observed clustering of the peak energy of the $E\cdot
f(E)$ spectrum in the sub-MeV range and its quantitative comparison
with observations indicates that the combination of parameters 
$(\varepsilon \Omega)$ (where  $\varepsilon$ stands for the fraction of the Alfv\'en speed with which
reconnection takes place and $\Omega$ for the angular frequency of the rotator)
has values $(\varepsilon \Omega)\sim 100$ rad s$^{-1}$.
This is compatible with reconnection speed proceeding with
$\varepsilon \sim 0.03$ and a rotator with angular frequency $\Omega\sim
10^{3.5}$rad s$^{-1}$. The Amati relation (i.e. $E_{\rm p}\sim E_{\rm iso}^{0.5}$;
Amati et al. 2002; 2006) can be explained by the model provided that the most luminous bursts are 
characterized by higher energy-to-rest-mass ratio. 

More constraints to both the model and the operation of the central engine  
can come from the study of the variability properties of the GRBs.
In the photospheric model, the observed variability is a direct manifestation of
the activity of the central engine. Individual pulses, considered to be the
building blocks of the GRB lightcurves (e.g.  Fenimore et
al. 1995) come from modulations of the luminosity and the baryon loading
of the flow and are ultimately linked to the central engine variability.  
Modeling of the observed properties of the GRB pulses can, thus, provide
us with a unique probe into oparation of the central engine on diferent
timescales during the GRB prompt phase.   

We have simulated GRB pulses by assuming that the luminosity of the flow
varies following the typical observed pulse profile (cf. Norris et al. 1996)
and that the magnetization parameter $\sigma_0$ corelates with the luminosity 
of the flow as indicated by our interpretation of the Amati relation. The
pulse width predicted by the model is narrower in the harder 
energy bands, in agreement with observation.

\appendix
\section{Fitting functions for the spectral parameters}

Here we give the expressions that describe how the
photospheric luminosity $L_{\rm ph}$, the peak energy $E_{\rm p}$,
the high frequency power law $\lambda_2$ and the spectral break
$\Lambda$ depend on the physical parameters $L$, $\sigma_0$ and
$\varepsilon \Omega$ of the reconnection model. 
We first, derive some useful expressions related to the SBPL
model that will simplify the apearance of the fitting formulas.

\subsection{A note on the Smoothly connected power-law model}

The expression that describes the SBPL model can be quite
generally written (e.g., Ryde 1999)
\be
g(x)=A\Big(\frac{x}{x_{\rm piv}}\Big)^{\frac{\lambda_1+\lambda_2}{2}}
\Big(\frac{e^q+e^{-q}}{e^{q_{\rm piv}}+e^{-q_{\rm piv}}}\Big)^{\frac{\lambda_2-\lambda_1}{2}\Lambda},
\label{SBPLapp}
\ee
where
\be
q=\frac{\ln(x/x_{\rm b})}{\Lambda}, \quad q_{\rm piv}=\frac{\ln(x_{\rm piv}/x_{\rm b})}{\Lambda} \nonumber.
\ee
The value $x_{\rm piv}$ is a reference value of the $x$ variable one has the
freedom to choose. For $x=x_{\rm piv}$ one easily check that $g(x_{\rm piv})=A$. 

This expression can be simplified if the reference value $x_{\rm piv}$ is
taken sufficiently far from the break $x_{\rm b}$. If, for example, 
$e^{q_{\rm {piv}}}\gg e^{-q_{\rm {piv}}}$, one can simplify the
expression for the SBPL model.

Under the condition that $e^{q_{\rm {piv}}}\gg e^{-q_{\rm {piv}}}$  
eq.~(\ref{SBPLapp}) can be brought into the form
\be
g(x)=A\Big(\frac{x}{x_{\rm piv}}\Big)^{\frac{\lambda_1+\lambda_2}{2}}
\Big(\frac{x_{\rm b}}{x_{\rm piv}}\Big)^{\frac{\lambda_2-\lambda_1}{2}}
\Big(e^q+e^{-q}\Big)^{\frac{\lambda_2-\lambda_1}{2}\Lambda}.
\label{SBPLsim}
\ee
One can also derive the asymptotic expressions for the low and high 
power-law dependence of $g(x)$ on $x$ in the limits $e^q\gg e^{-q}$ and vice versa. 
So expression (\ref{SBPLsim}) gives
\be
g(x)=A\Big(\frac{x}{x_{\rm piv}}\Big)^{\lambda_2}, \quad e^q\gg e^{-q} 
\label{highslope}
\ee and
\be g(x)=A\Big(\frac{x}{x_{\rm piv}}\Big)^{\lambda_1}
\Big(\frac{x_{\rm b}}{x_{\rm piv}}\Big)^{\lambda_2-\lambda_1}, 
\quad e^q\ll e^{-q}.
\ee
The form (\ref{SBPLsim}) of the SBPL model is used below for fitting the 
dependency of the  photospheric spectrum on the parameters of our
magnetically powered GRB flow.

\subsection{Fitting formulas}

The dependency of the photospheric bolometric luminosity on $\sigma_0$
is well modeled by a SBPL model (see fig.~3) of the form
\be
L_{\rm ph}=A_{\rm L}
\Big(\frac{\sigma_0}{100}\Big)^{1.5}\Big(\frac{\sigma_{\rm
 0,cr}}{100}\Big)^{-3}\Big(e^{q}+e^{-q}\Big)^{-0.5},
\label{SBPLlum}
\ee
where
\be
q=6\ln(\sigma_0/\sigma_{\rm 0,cr}) \nonumber
\ee   
and $\sigma_{\rm 0,cr}=42 \big(L_{52}(\varepsilon \Omega)_3\big)^{2/15}$ (defined in eq.~(\ref{sigmacr})).
The best fit for $A_{\rm L}$  is
\be
A_{\rm L}\approx 7\cdot10^{50}(\varepsilon\Omega)_3^{0.2}L_{52}^{1.2}\quad
\rm{erg\cdot s^{-1}}. \label{lumparameters}
\ee

The rather sharp break means that for $\sigma_0>\sigma_{\rm 0,cr}$
the photospheric luminosity is well represented by (see also
eq.~(\ref{highslope}))
\be
L_{\rm  ph}=7\cdot10^{50}(\varepsilon\Omega)_3^{0.2}L_{52}^{1.2}\sigma_{0,2}^{-1.5}\quad
\rm{erg\cdot s^{-1}}. \qquad (\sigma>\sigma_0)
\label{Lhigh}
\ee

The  fit for the peak energy of the $E\cdot f(E)$ spectrum on $\sigma_0$ is 
\be
E_{\rm p}=A_{\rm p}
\Big(\frac{\sigma_0}{100}\Big)^{3.6}\Big(\frac{\sigma_{\rm
 0,cr}}{100}\Big)^{-3.3}\Big(e^q+e^{-q}\Big)^{-0.55},
\label{SBPLep}
\ee
with
\be
A_{\rm p}\approx 1200(\varepsilon\Omega)_3^{0.33}L_{52}^{0.11}\quad {\rm keV}. 
\label{epparameters}
\ee

For $\sigma_0>\sigma_{\rm 0,cr}$ this reduces to (cf. eq.~(\ref{highslope}))
\be
E_{\rm p}\approx 1200(\varepsilon\Omega)_3^{0.33}L_{52}^{0.11}\sigma_{0,2}^{0.3}\quad {\rm keV}.
\label{Ephigh}
\ee

The fit for the high-energy slope is
\be
\lambda_2\approx-2.5
\Big(\frac{\sigma_0}{100}\Big)^{-1.8}\Big(\frac{\sigma_{\rm
 0,cr}}{100}\Big)^{1.8}\Big(e^q+e^{-q}\Big)^{0.3}.
\label{SBPLep}
\ee

The break scale is approximately constant, $\Lambda\simeq 1.6$,
for high $\sigma_0$ and makes a transition to $\Lambda\simeq 2.8$
for low values of  $\sigma_0$. The transition is modeled with the hyperbolic
tangent function with best fit 
\be
\Lambda\approx2.2-0.6\tanh\Big(8\frac{\sigma_0-\sigma_{0,{\rm cr}}}{\sigma_{0,{\rm cr}}}\Big).
\label{elparameters}
\ee
 
An approximation to the spectrum itself, as a function
of baryon loading, flow luminosity and reconnection rate, is given by 
eq.~(\ref{SBPL}), with parameters given by the fitting formulas above. The
normalization factor $A$ in eq.~(\ref{SBPL}) follows by equating the
integrated luminosity $\int f(E)\, {\rm d}E$ to the photospheric luminosity
from \ref{SBPLlum}.


\begin{thebibliography}{}

\bibitem{} Abramowicz, M. A., Novikov, I. D., \& Paczy\'nski, B. 1991, ApJ,
  369, 175
\bibitem{} Amati, L., Frontera, F., Tavani, M., et al. 2002, A\&A, 390, 81
\bibitem{} Amati, L. 2006, MNRAS, 372, 233
\bibitem{} Band, D., \& Preece, R. D. 2005, ApJ, 627, 319
\bibitem{} Band, D., Matteson, J., Ford, L., et al. 1993, ApJ, 413, 281
\bibitem{} Blandford, R. D., \& Payne, D. G. 1982, MNRAS, 199, 883
\bibitem{} Coroniti, F. V. 1990, ApJ, 349, 538
\bibitem{} Crider, A., Liang, E. P., Smith, I. A., et al. 1997, ApJ, 479, L39
\bibitem{} Daigne, F., \& Mochkovitch, R. 2002, MNRAS, 336, 1271 
\bibitem{} Drenkhahn, G. 2002, A\&A, 387, 714
\bibitem{} Drenkhahn, G., \& Spruit, H. C. 2002, A\&A, 391, 1141
\bibitem{} Fenimore, E. E., in 't Zand, J. J. M., Norris, J. P., Bonnell,
  J. T., \& Nemiroff, R. J. 1995, ApJ, 448, L101
\bibitem{} Frontera, F., Amati, L., Costa, E., et al. 2000, ApJS, 127, 59
\bibitem{} Ghirlanda, G., Celotti, A., \& Ghisellini, G. 2003, A\&A, 406, 879
\bibitem{} Ghisellini, G., \& Celotti, A. 1999, A\&AS, 138, 527
\bibitem{} Giannios, D. 2006, A\&A, 457, 763
\bibitem{} Giannios, D., \& Spruit, H. C. 2005, A\&A, 430, 1
\bibitem{} Giannios, D., \& Spruit, H. C. 2006, A\&A, 450, 887 
\bibitem{} Goodman, J. 1986, ApJ, 308, L47
\bibitem{} Kaneko, Y., Preece, R. D., Briggs, M. S., et al. 2006, ApJS, 166, 298
\bibitem{} Lyubarsky, Y. E. 2005, MNRAS, 358, 113
\bibitem{} Lyutikov, M., \& Blandford R. D. 2003, astro-ph/0312347
\bibitem{} M\'esz\'aros, P., \& Rees, M. J. 1997, ApJ, 482, L29
\bibitem{} M\'esz\'aros, P., \& Rees, M. J. 2000, ApJ, 530, 292
\bibitem{} Michel, F. C. 1969, ApJ, 158, 727
\bibitem{} Nakar, E., \& Piran, T. 2005, MNRAS, 360, L73
\bibitem{} Norris, J. P., Nemiroff, R. J., Bonnell, J. T., et al. 1996, ApJ,
 459, 393
\bibitem{} Paczy\'nski, B. 1986, ApJ, 308, L43
\bibitem{} Pe'er, A., M\'esz\'aros, P., \& Rees, M. J. 2005, ApJ, 635, 476
\bibitem{} Piran, T. 2005, Rev. Mod. Phys., 76, 1143
\bibitem{} Piran, T., Shemi, A., \& Narayan, R. 1993, MNRAS, 263, 861
\bibitem{} Preece, R. D., Briggs, M. S., Mallozzi, R. S, \& Brock, M. N. 1994,
  ``WINdows Gamma SPectral ANalysis (WINGSPAN)''
\bibitem{} Preece, R. D., Briggs, M. S., Mallozzi, R. S, et al. 1998, ApJ,
  506, L23
\bibitem{} Ramirez-Ruiz, E. 2005, MNRAS, 363, L61
\bibitem{} Rees, M. J., \& M\'esz\'aros, P. 1994, ApJ, 430, L93
\bibitem{} Ryde, F. 1999, Astroph. Lett. and Comm., 39, 281
\bibitem{} Ryde, F. 2004, ApJ, 614, 827
\bibitem{} Ryde, F. 2005, ApJ, 625, L95
\bibitem{} Sari, R., \& Piran, T. 1997, MNRAS, 287, 110
\bibitem{} Spruit, H. C., \& Drenkhahn, G. D. 2004, Magnetically powered
  prompt radiation and flow acceleration in GRB, in Proceedings ``Gamma Ray
  Bursts in the Afterglow Era, Third Workshop'', Rome, Sept 2002. ASP
  Conference Proccedings, 312, 357
\bibitem{} Spruit, H. C., Daigne, F., \& Drenkhahn, G. 2001, A\&A, 369, 694
\bibitem{} Thompson, C. 1994, MNRAS, 270, 480
\bibitem{} Thompson, C., M\'esz\'aros, P., \& Rees, M. J. 2006, ApJ,
  submitted, astro-ph/0608282
\bibitem{} Uzdensky, D. A., \& MacFadyen, A. I. 2006, ApJ, 647, 1192 
\bibitem{} Vlahakis, N., \& K\"onigl, A. 2003, ApJ, 596, 1080

\end{thebibliography}
\end{document}